\documentstyle[twocolumn,aps]{revtex}

\def\go{\mathrel{\raise.3ex\hbox{$>$}\mkern-14mu
             \lower0.6ex\hbox{$\sim$}}}
\def\lo{\mathrel{\raise.3ex\hbox{$<$}\mkern-14mu
             \lower0.6ex\hbox{$\sim$}}}
\def\eps{{\varepsilon}}

\begin{document}

\title{Superfluidity of Hydrogenlike Gas in a Strong Magnetic Field ?}

\author{Dong Lai}
\address{Theoretical Astrophysics, 130-33, California Institute of Technology,
Pasadena, CA 91125\\
}
\date{March 1995}
\maketitle

{\bf Abstract:\/}
The recent claim that in a strong magnetic field hydrogenlike gas
(e.g., excitons in certain semiconductors, neutron star surface layers)
becomes superfluid is refuted. Molecules form by strong covalent bond
along the magnetic field axis, which prohibits Bose-Einstein
condensation.

\bigskip
\centerline{To appear in {\it Phys. Rev. Lett.} (1995) as a comment}
\bigskip

\hrule
\bigskip
\bigskip

In a recent letter and several papers Korolev and Liberman
\cite{Korolev94,Korolev93} claimed
that in a strong magnetic field a hydrogenlike gas
(e.g., excitons in certain semiconductors) can form
Bose-Einstein condensate and become a superfluid. This is an
interesting result as it potentially provides another system besides
liquid He that displays Bose-Einstein condensation.
However, the authors\cite{Korolev94,Korolev93}
failed to identify the strong covalent bonding
mechanism for forming hydrogen molecule in strong magnetic
field\cite{Lai92,Ruderman74}, which makes the Bose-Einstein
condensation rather unlikely.

For definiteness, I consider the electron-proton system.
For excitons in semiconductors, the results (to the leading order)
can be rescaled by introducing the effective electron mass
and the dielectric constant of the medium.

The atomic unit (a.u.) for the magnetic field strength is
$B_o=m_e^2e^3 c/\hbar^3=2.35 \times 10^9~{\rm G}$, above which the
Landau excitation energy of the electron $\hbar\omega_e=\hbar eB/(m_ec)$
is larger than the atomic unit for the energy $m_ee^4/\hbar^2=2$ Ry.
In the strong field regime $b\equiv B/B_o>>1$,
the electron settles into the ground Landau state, and
the energy levels of the atom are specified by two
quantum numbers $(m,\nu)$: $m$ measures the mean transverse
separation between the electron guiding center and the proton,
$\rho_m=(2m+1)^{1/2}\hat\rho$ ($m=0,1,2,\cdots$),
where $\hat\rho=(\hbar c/eB)^{1/2}=b^{-1/2}$ (a.u.),
and $\nu$ is the number of nodes in the $z$-wavefunction of the electron.
The $\nu\neq 0$ states have binding energies less than a Rydberg.
For the $(m,\nu)=(m,0)$ state, the atom is elongated along magnetic
field axis, and the energy is approximately given by
(see\cite{Lai92} and references therein)
$$E_m\simeq -0.16 \left(\ln{b\over 2m+1}\right)^2~({\rm a.u.}),
$$
for $b\sim 10-10^{5}$ (Incidentally, Refs.\cite{Korolev94,Korolev93}
incorrectly assume the coefficient 
to be $0.5$, which can only be approached asymptotically
as $b\rightarrow\infty$).
Thus the ground-state binding energy $|E_0|\simeq 0.16 (\ln b)^2$ can be
much larger than Rydberg when $b>>1$.

Since the electron spins in the atoms are all aligned
antiparallel to the magnetic field, two atoms in their
ground states ($m=\nu=0$) cannot easily bind together to form a
molecule according to the exclusion principle. Approximate
calculations\cite{Korolev93} indicate that the interaction
between two ground-state hydrogen atoms is very weak
(However, this calculation underestimates the binding energy
because it neglects the overlapping of the electron wavefunctions.
Recent calculations\cite{Lai95} indicate that
the binding energy  
is much larger than the result of\cite{Korolev93}).
This formed the basis of the claim in\cite{Korolev94}
that hydrogenlike gas can become superfluid: because of the weak interatomic
interaction, the system remains in the gaseous phase even at low
temperature and high density, hence Bose-Einstein condensation may
eventually sets in as temperature decreases or density increases.

However, these is another mechanism by which
tightly-bound hydrogen molecule can form in strong
magnetic field\cite{Lai92,Ruderman74}:
One of the atom is first excited to a $m>0$ state;
then the two atoms, one in the ground $m=0$
state, another in the $m=1$ state, can easily form a spin-triplet molecule
by covalent bonding along the magnetic field direction.
Since the ``activation energy'' for exciting an electron in the atom from
Landau orbital $m$ to $(m+1)$ is small (of order $\ln b$), the
covalent binding energy more than compensates for the ``activation energy'',
and the resulting molecule can be tightly bound and stable.
In this way, more atoms can be added along the magnetic field axis to
form a molecular chain.
Detailed Hartree-Fock calculations of the electronic structure of
hydrogen poly-molecules have been performed\cite{Lai92} for the strong field
regime $b>>1$. The dissociation energy of H$_2$
molecule approximately scales as
$D\sim 0.03(\ln b)^2$ (a.u.), while the equilibrium interatomic separation
scales as $1/\ln b$ (the numerical results can be found in\cite{Lai92}).
For example, $D\simeq 0.3$ (a.u.) at $b=42.5$,
and $D\simeq 0.6$ (a.u.) at $b=10^2$.
This is more than two orders of magnitude larger than the interaction
potential depth (about $10^{-3}$ a.u. at $b=10^2$)
discussed in\cite{Korolev93}.

Because of the strong tendency of forming molecules,
a hydrogenlike gas in strong magnetic field can not behave as a
weakly-interacting boson gas, therefore the
conclusions of\cite{Korolev94,Korolev93} are invalid, at least for $b>>1$
(For $b$ just slightly larger than unity, no reliable calculation
of the H molecule has been performed so far).
As the gas density increases or temperature decreases,
more molecules and chains will form. Eventually, because of the
strong chain-chain interaction, the system may transform into
a metallic state\cite{Lai95}.

We note that in the laboratory field regime, the condensation
of spin-polarized hydrogen is a real possibility,
although not yet achieved\cite{Silvera92,Walraven94}.

Finally, it is interesting to speculate the possible detectable
effects of the molecules formed from two electron-hole excitons
in semiconductors. If we adopt the typical values for
the effective electron mass $m_{\rm eff}\sim 0.01m_e$ and the
dielectric constant $\eps\sim 10$, then the unit for the energy is
$2 Ry'=m_{\rm eff}e^4/(\eps^2\hbar^2)\sim 3\times 10^{-3}$ eV and the unit for
the magnetic field is $B_c'=m_{\rm eff}^2e^3c/(\eps^2\hbar^3)
\sim 2\times 10^3$ G.
Therefore for $B=20$ Tesla ($b=B/B_c'=100$), the exciton binding
energy is $\sim 10^{-2}$ eV, while the dissociation energy
for the exciton molecule is $\sim 2\times 10^{-3}$ eV.


\end{document}